\documentclass[journal = ancac3]{achemso}
\setkeys{acs}{usetitle = true}

\usepackage{graphicx}
\usepackage{color}
\usepackage{amsfonts}

\usepackage{isomath}
\usepackage{amsmath}
\usepackage{amsthm}
\usepackage{amsfonts}
\usepackage{amssymb}
\usepackage{fdsymbol}
\usepackage{braket}

\usepackage{multirow}
\usepackage[table,xcdraw]{xcolor}

\let\oldAA\AA
\renewcommand{\AA}{\text{\normalfont\oldAA}}

\title{\vspace{-1.5cm}Visualizing band structure hybridization and superlattice effects in twisted MoS$_2$/WS$_2$ heterobilayers\vspace{-0.2cm}}

\author{Alfred J. H. Jones}
\affiliation{Department of Physics and Astronomy, Aarhus University, 8000 Aarhus C, Denmark}
\altaffiliation{Contributed equally to this work}
\author{Ryan Muzzio}
\affiliation{Department of Physics, Carnegie Mellon University, Pittsburgh, Pennsylvania 15213, USA}
\altaffiliation{Contributed equally to this work}
\author{Sahar Pakdel}
\affiliation{Department of Physics and Astronomy, Aarhus University, 8000 Aarhus C, Denmark}
\alsoaffiliation{CAMD, Department of Physics, Technical University of Denmark, 2800 Kgs. Lyngby, Denmark}
\author{Deepnarayan Biswas}
\author{Davide Curcio}
\author{Nicola Lanat\`a}
\affiliation{Department of Physics and Astronomy, Aarhus University, 8000 Aarhus C, Denmark}
\alsoaffiliation{Nordita, KTH Royal Institute of Technology and Stockholm University, Roslagstullsbacken 23, 10691 Stockholm, Sweden}
\author{Philip~Hofmann}
\affiliation{Department of Physics and Astronomy, Aarhus University, 8000 Aarhus C, Denmark}
\author{ Kathleen M. McCreary}
\author{ Berend T. Jonker}
\affiliation{Naval Research laboratory, Washington, D.C. 20375, USA}
\author{Kenji~Watanabe}
\affiliation{Research Center for Functional Materials, 
National Institute for Materials Science, 1-1 Namiki, Tsukuba 305-0044, Japan}
\author{Takashi~Taniguchi}
\affiliation{International Center for Materials Nanoarchitectonics, 
National Institute for Materials Science,  1-1 Namiki, Tsukuba 305-0044, Japan}
\author{Simranjeet Singh}
\affiliation{Department of Physics, Carnegie Mellon University, Pittsburgh, Pennsylvania 15213, USA}
\author{Roland J. Koch}
\affiliation{Advanced Light Source, E. O. Lawrence Berkeley National Laboratory, Berkeley, California 94720, USA}
\author{Chris Jozwiak}
\affiliation{Advanced Light Source, E. O. Lawrence Berkeley National Laboratory, Berkeley, California 94720, USA}
\author{Eli Rotenberg}
\affiliation{Advanced Light Source, E. O. Lawrence Berkeley National Laboratory, Berkeley, California 94720, USA}
\author{Aaron Bostwick}
\affiliation{Advanced Light Source, E. O. Lawrence Berkeley National Laboratory, Berkeley, California 94720, USA}
\author{Jill A. Miwa}
\affiliation{Department of Physics and Astronomy, Aarhus University, 8000 Aarhus C, Denmark}
\author{Jyoti Katoch}
\email{jkatoch@andrew.cmu.edu}
\affiliation{Department of Physics, Carnegie Mellon University, Pittsburgh, Pennsylvania 15213, USA}
\author{S{\o}ren Ulstrup\vspace{-0.2cm}}
\email{ulstrup@phys.au.dk}
\affiliation{Department of Physics and Astronomy, Aarhus University, 8000 Aarhus C, Denmark}
\begin{document}
\newpage
\begin{abstract}

A mismatch of atomic registries between single-layer transition metal dichalcogenides (TMDs) in a two dimensional van der Waals heterostructure produces a moir\'e superlattice with a periodic potential, which can be fine-tuned by introducing a twist angle between the materials. This approach is promising both for controlling the interactions between the TMDs and for engineering their electronic band structures, yet direct observation of the changes to the electronic structure introduced with varying twist angle has so far been missing. Here, we probe heterobilayers comprised of single-layer MoS$_2$ and WS$_2$ with twist angles of $(2.0 \pm 0.5)^{\circ}$, $(13.0 \pm 0.5)^{\circ}$, and $(20.0 \pm 0.5)^{\circ}$ and investigate the differences in their electronic band structure using micro-focused angle-resolved photoemission spectroscopy. We find strong interlayer hybridization between MoS$_2$ and WS$_2$ electronic states at the $\bar{\mathrm{\Gamma}}$-point of the Brillouin zone, leading to a transition from a direct bandgap in the single-layer to an indirect gap in the heterostructure. Replicas of the hybridized states are observed at the centre of twist angle-dependent moir\'e mini Brillouin zones. We confirm that these replica features arise from the inherent moir\'e potential by comparing our experimental observations with density functional theory calculations of the superlattice dispersion. Our direct visualization of these features underscores the potential of using twisted heterobilayer semiconductors to engineer hybrid electronic states and superlattices that alter the electronic and optical properties of 2D heterostructures. 

\end{abstract}

\newpage

Stacking of two dimensional (2D) van der Waals (vdW) materials is a promising avenue for tailored optoelectronic applications and engineering electronic states of semiconductor devices\cite{Geim2013,Xu2013,Jariwala2014}. Despite relatively weak out of plane interactions between single layer (SL) vdW materials, their electronic properties may be dramatically altered via vertical stacking into homo- or hetero-bilayers. Forming heterobilayers of semiconducting transition metal dichalcogenides (TMDs) such as MoS$_2$ and WS$_2$ has proven a viable route to engineer type II (staggered gap) semiconductor band alignments, which can be utilized for ultrafast charge transfer and spatial separation of light-induced electron-hole pairs, ideal for energy harvesting \cite{Hong2014,Schaibley2016,Chen2016, Deilmann2018}. Furthermore, these materials feature strong spin-orbit coupling and many-body interactions that lead to tightly bound excitons, which inherit the valley pseudospin of the constituent electrons and holes, making the materials highly promising candidates for valleytronic applications \cite{Xiao2010,Li:2014,Rivera2018_interlayerexcitonreview}.

These remarkable properties of TMD heterostructures can be further modified by introducing a moir\'e superlattice via the mismatch of the atomic registries of the TMD single layers, which can be controllably tuned by varying the twist angle, $\theta$, between layers. This moir\'e potential folds the bands into mini Brillouin zones (mBZ) where the new superlattice bands hybridize. At an optimal value of $\theta$ these hybrid states may trigger emergent phenomena, as evidenced by the appearance of competing superconducting and correlated insulating charge orders in twisted bilayer graphene \cite{Lopes2007,Cao2018,Cao:2018b,Zhang2020}. Such phenomena have also been predicted to emerge in homo- and heterobilayers of twisted TMDs \cite{Ruiz-Tijerina2019,Naik2018,Wang2020}. In these heterostructures, it is also possible to confine excitons within the superlattice potential\cite{Wu2017,Wu2018,Seyler2019, Tran2019,Alexeev2019} and utilize the twist angle as a tuning knob for the exciton lifetime \cite{Choi2021}.

The real-space moir\'e superlattice potential, the associated electronic density of states, and excitonic effects resulting from band hybridization in twisted TMD heterostructures have been investigated using  scanning probe microscopies \cite{Hill2016,Chen2019} and all-optical approaches \cite{Zande2014,Wang2016_WSe2/WS2,Alexeev2019,Seyler2019}. Band alignments in TMD heterostructures have been measured using photoemission spectroscopies \cite{Jin2015,Chiu2015,Wilson2017, Ulstrup2019_nanoWS2}. However, systematic, direct studies of the energy- and momentum-dependent electronic band structure of TMD vdW heterostructures subject to $\theta$-dependent superlattices have not been reported, leaving open questions on the dispersion of the superlattice minibands and hybridization effects in these materials. 

Here, we have devised a strategy to close this knowledge gap, by transferring multiple microscopic SL MoS$_2$ islands onto an extended flake of SL WS$_2$ supported on hexagonal boron nitride (hBN), providing access to multiple twisted TMD heterostructures with $\theta =\{(2.0 \pm 0.5)^{\circ},(13.0 \pm 0.5)^{\circ},(20.0 \pm 0.5)^{\circ}\}$ within the same sample. We then utilize angle resolved photoemission spectroscopy with a micro-focused synchrotron beam (microARPES) based on a capillary focusing approach \cite{Koch2018,Ulstrup2020} to measure the electronic structure of each heterostructure with a spatial resolution of $(1.8 \pm 0.3)$~$\mu$m. Our measurements reveal strong hybridization effects between MoS$_2$ and WS$_2$ bands at the $\bar{\mathrm{\Gamma}}$-point of the BZ, characterized by a strong energy splitting of $(548 \pm 33)$~meV and minibands around the center of mBZs, i.e. $\bar{\mathrm{\Gamma}}_m$. These interactions lead to a transition from a direct bandgap in the SL TMDs to an indirect bandgap in our heterostructures. We compare our experimental dispersion with density functional theory (DFT) calculations of unfolded superlattice electronic bands in relaxed MoS$_2$/WS$_2$ structures with commensurate stacking near our experimental twist angles. We find agreement between experimental and theoretical bands, which strongly suggests that the measured minibands arise from the inherent moir\'e potential. The calculations provide an estimate for the conduction band dispersion and, in combination with the measured valence band dispersion, reveal new indirect transitions caused by the superlattice.
 
\section{Results and Discussion}

\begin{figure*} [t!]
	\begin{center}
		\includegraphics[width=1\textwidth]{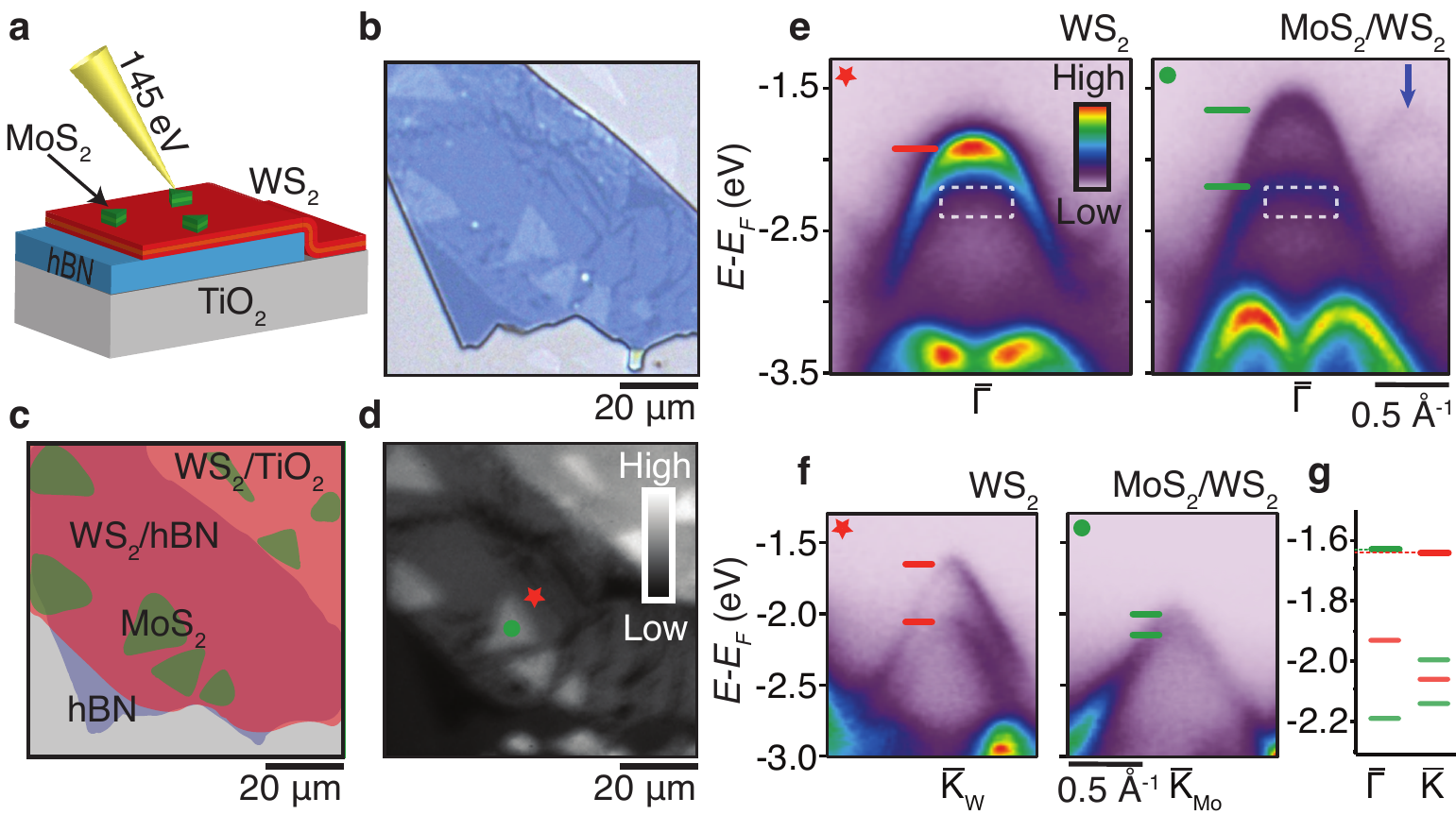}
		\caption{Micro-focused photoemission experiment on multiple MoS$_2$/WS$_2$ heterostructures: (a) Sketch of sample layout: A SL WS$_2$ flake (red) sits on a thick hBN island (blue) and a TiO$_2$ substrate (grey). Multiple microscopic islands of differently oriented SL MoS$_2$ (green) have been transferred on top of the SL WS$_2$ flake. The yellow cone illustrates the micro-focused beam of photons utilized for microARPES. (b) Optical image of the sample. The contrast has been digitally enhanced to highlight SL flakes. (c) Simplified schematic and identification of main features from (b). (d) Map of $(E,k)$-integrated photoemission intensity measured over the same area as seen in (b)-(c). (e)-(f) ARPES $E(k)$-dispersion measured around (e) $\bar{\mathrm{\Gamma}}$ and (f) $\bar{\mathrm{K}}_{\mathrm{W}}$ for the areas of the SL WS$_2$ (left column) and $\bar{\mathrm{K}}_{\mathrm{Mo}}$ for the areas of MoS$_2$/WS$_2$ (right column) marked by corresponding symbols in (d). Red and green tick marks denote local band maxima extracted from energy distribution curves. The blue arrow in the MoS$_2$/WS$_2$ dispersion in (e) marks a moir\'e replica band. The white dashed box in (e) demarcates the $(E,k)$-region of intensity, which is integrated to produce the map in (d). (g) Summary of band maxima at $\bar{\mathrm{\Gamma}}$ and $\bar{\mathrm{K}}$ for SL WS$_2$ (red ticks) and MoS$_2$/WS$_2$ (green ticks). The different VBM between the two regions has been highlighted with thin dashed lines. The MoS$_2$/WS$_2$ heterostructure shown here has a twist angle of 20$^\circ$. The ARPES data here was acquired at 145 eV photon energy. }
		\label{fig:1}
	\end{center}
\end{figure*}

A schematic of our photoemission experiment and the composition of our sample is presented in Figure \ref{fig:1}(a). SL MoS$_2$ islands with different orientations are transferred on a SL WS$_2$ flake, which straddles a thick hBN island and a supporting conductive Nb-doped TiO$_2$ substrate. The hBN provides a flat and inert support for the SL TMDs, while the doped TiO$_2$ ensures that all heterostructure components are well-grounded, such that charging effects are avoided and high quality photoemission spectra can be collected \cite{katoch2018,ulstrup2019xs}.  An optical image of the sample is presented in Figure \ref{fig:1}(b). The major blue structure corresponds to the hBN island while the increasingly lighter hues of blue indicate SL WS$_2$ and triangular shaped SL MoS$_2$ flakes. The main features and their shapes are illustrated in Figure \ref{fig:1}(c) as a general guideline. We directly compare the optical image in Figure \ref{fig:1}(b) with a map of $(E,k)$-integrated photoemission intensity over the same area, as shown in Figure \ref{fig:1}(d). Such a map is obtained by collecting a 2D ARPES snapshot at each position on the sample, leading to a four-dimensional (4D) dataset containing the $(E,k,x,y)$-dependent photoemission intensity. This enables us to correlate the real-space $(x,y)$-dependent features of our sample with the ``local" $E(k)$-dispersion \cite{katoch2018,Koch:2018ab,Kastl2019}. This method leads to our observation of the SL WS$_2$ and MoS$_2$/WS$_2$ bandstructures around $\bar{\mathrm{\Gamma}}$ and the $\bar{\mathrm{K}}$-points for both MoS$_2$ and WS$_2$, which we label as $\bar{\mathrm{K}}_{\mathrm{Mo}}$ and $\bar{\mathrm{K}}_{\mathrm{W}}$, respectively. The cuts considered in Figures \ref{fig:1}(e)-(f) have been extracted from 3D $(E,k_x,k_y)$ data acquired from the areas marked by a green circle and a red star in Figure \ref{fig:1}(d). The $(x,y)$-dependent intensity in Figure \ref{fig:1}(d) has been obtained by integrating the $(E,k)$-intensity from the same area as shown by a white dashed box in Figure \ref{fig:1}(e), providing strong contrast between MoS$_2$/WS$_2$ (see white triangular structures) and bare WS$_2$ regions.

The SL WS$_2$ ARPES dispersion seen in Figures \ref{fig:1}(e)-(f), for the $\bar{\mathrm{\Gamma}}$ (Figure \ref{fig:1}(e)) and $\bar{\mathrm{K}}_{\mathrm{W}}$ (Figure \ref{fig:1}(f)) points respectively, presents a sharp hole-like band with a local maximum at $\bar{\mathrm{\Gamma}}$ and two spin-orbit split branches with a global maximum at $\bar{\mathrm{K}}_{\mathrm{W}}$. The splitting, as determined from energy distribution curve (EDC) analysis presented in Figure 7 of the supporting information, is found to be $(432 \pm 31)$~meV. The measured splitting and band alignments are in agreement with previous ARPES measurements on SL WS$_2$ transferred on hBN \cite{katoch2018,ulstrup2019xs} and DFT calculations of free-standing SL WS$_2$ \cite{Zhu2011}, indicating the high quality and quasi-free-standing nature of the SL WS$_2$ on hBN. By comparison, the ARPES dispersion from the MoS$_2$/WS$_2$ heterostructure shows that the valence band (VB) states at $\bar{\mathrm{\Gamma}}$ have split into two bands, separated by $\Delta E_{\bar{\mathrm{\Gamma}}} = (548 \pm 33)$~meV. At $\bar{\mathrm{K}}_{\mathrm{Mo}}$, the states do not exhibit a clear splitting, as was observed in the SL WS$_2$, but instead appear significantly broadened with the expected splitting of 145 meV for SL MoS$_2$ unable to be resolved\cite{Zhu2011,miwa:2015}. The EDC fit at $\bar{\mathrm{K}}_{\mathrm{Mo}}$ was constrained to have a fixed splitting of 145 meV, which provides a satisfactory description of the measured spectrum. The EDC fits used to calculate the band positions are presented in Supporting Figure 7. 

In the MoS$_2$/WS$_2$ heterostructure, the additional splitting at $\bar{\mathrm{\Gamma}}$ resembles the bonding/antibonding splitting that has been observed in TMD homobilayers, which is a result of hybridization between the out-of-plane orbitals at $\bar{\mathrm{\Gamma}}$ between the two layers \cite{Jin2013,Kosmider2013,He2014,Signe:2015,Yeh2016, Wilson2017}. The presence of this splitting in our heterostructures is a clear indication of a significant interaction between the SL MoS$_2$ and SL WS$_2$, as we do not observe a simple superposition of the two SL band structures. Intriguingly, an additional hole-like spectral feature, labeled with a blue arrow to the right of $\bar{\mathrm{\Gamma}}$ in Figure \ref{fig:1}(e), indicates the replication of the hybridized bands at $\bar{\mathrm{\Gamma}}$. We will later confirm the $k$-position of these bands to be consistent with the moir\'e band structure unfolded onto the SL MoS$_2$ BZ, and show that this replica band is in  fact degenerate with the upper VB at $\bar{\mathrm{\Gamma}}$ in the Supplementary Figure 8. It is an important consequence of the new bands at $\bar{\mathrm{\Gamma}}$ that the global VB maximum shifts from $\bar{\mathrm{K}}_{\mathrm{W}}$ in the SL WS$_2$ to $\bar{\mathrm{\Gamma}}$ in MoS$_2$/WS$_2$, as seen in the diagram of the band positions at these two high symmetry points in Figure \ref{fig:1}(g) for SL WS$_2$ (red) and MoS$_2$/WS$_2$ (green). This shift is expected to cause a transition from the direct bandgap in SL WS$_2$ to an indirect bandgap in MoS$_2$/WS$_2$. As we will show later, the moir\'e replicas of these states will introduce the possibility of tunable optical transitions in the heterostructures.

\begin{figure*} [t!]
	\begin{center}
		\includegraphics[width=1\textwidth]{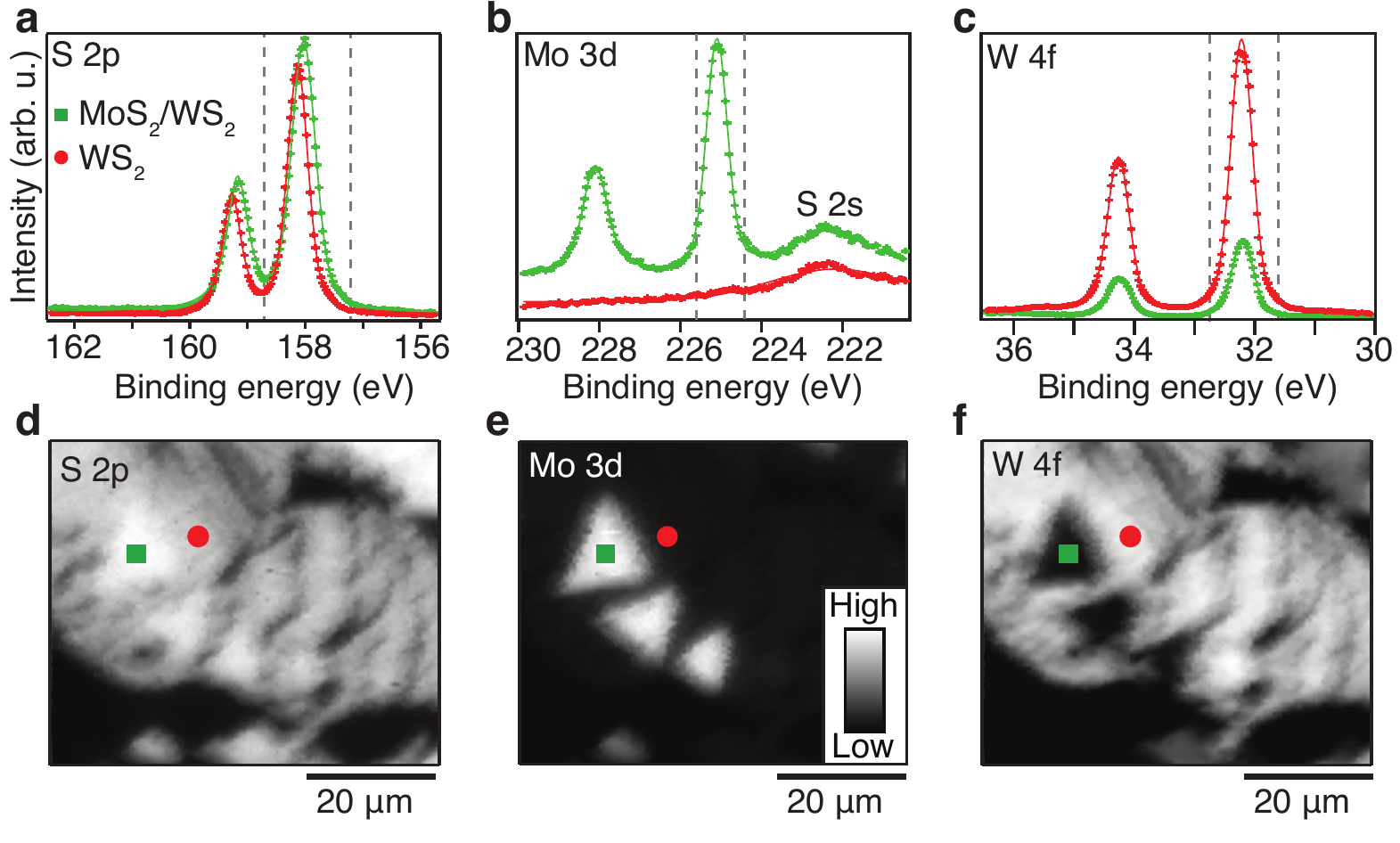}
		\caption{Spatially-resolved core level measurements: (a)-(c) Core level spectra corresponding to (a) S 2p, (b) Mo 3d and (c) W 4f binding energy regions obtained from areas with SL WS$_2$ (red circle) and a MoS$_2$/WS$_2$ heterostructure (green square). The measured core level spectra (markers) are overlaid with fits made using Voigt functions (smooth curves) (see Supporting Table 1 for fitting parameters). (d)-(f) Spatial dependence of core level intensity energy-integrated within the dashed grey vertical lines in the corresponding column of (a)-(c). The green square and red circle mark the areas where the core level spectra in (a)-(c) were obtained. The S 2p and Mo 3d core level spectra and maps were obtained using 350 eV photon energy, and the W 4f using 145 eV. }
		\label{fig:2}
	\end{center}
\end{figure*}

In order to unambiguously delineate areas with MoS$_2$/WS$_2$ and bare WS$_2$ in our sample, we examine the $(x,y)$-dependent intensity arising from the S 2p, Mo 3d and W 4f core level binding energy regions in Figure \ref{fig:2}. Example core level spectra from two such areas are shown in Figures \ref{fig:2}(a)-(c), with fitting parameters for these presented in Table 1 of the Supporting Information. Real space maps composed from the intensity integrated over the binding energy regions within the dashed vertical lines in Figure \ref{fig:2}(a)-(c), covering the corresponding core levels, are presented in Figures \ref{fig:2}(d)-(f). A minor chemical shift of 100 meV towards lower binding energies is seen in the S 2p binding energy region in Figure \ref{fig:2}(a) on the MoS$_2$/WS$_2$ area compared to the SL WS$_2$ area, however the overall features are very similar. This S 2p core level shift is expected, as the surface sensitive photoemission measurement primarily probes the Mo-S interaction on the MoS$_2$ islands instead of the W-S interaction on bare WS$_2$. The corresponding map in Figure \ref{fig:2}(d) exhibits a high intensity in areas with both MoS$_2$/WS$_2$  and bare WS$_2$, while the Mo 3d and W 4f binding energy regions in Figures \ref{fig:2}(b)-(c) present more striking differences between the two areas. The absence of Mo 3d peaks on bare WS$_2$, seen in Figure  \ref{fig:2}(b), leads to clear contrast from three triangular MoS$_2$ islands in Figure \ref{fig:2}(e). This contrast is inverted in the W 4f binding energy region in Figure \ref{fig:2}(f). However, W 4f core levels are also present on the triangular MoS$_2$ islands, as seen in Figure \ref{fig:2}(c), which ascertains that intact WS$_2$ is present under the MoS$_2$ flakes.

\begin{figure*} [t!]
	\begin{center}
		\includegraphics[width=0.7\textwidth]{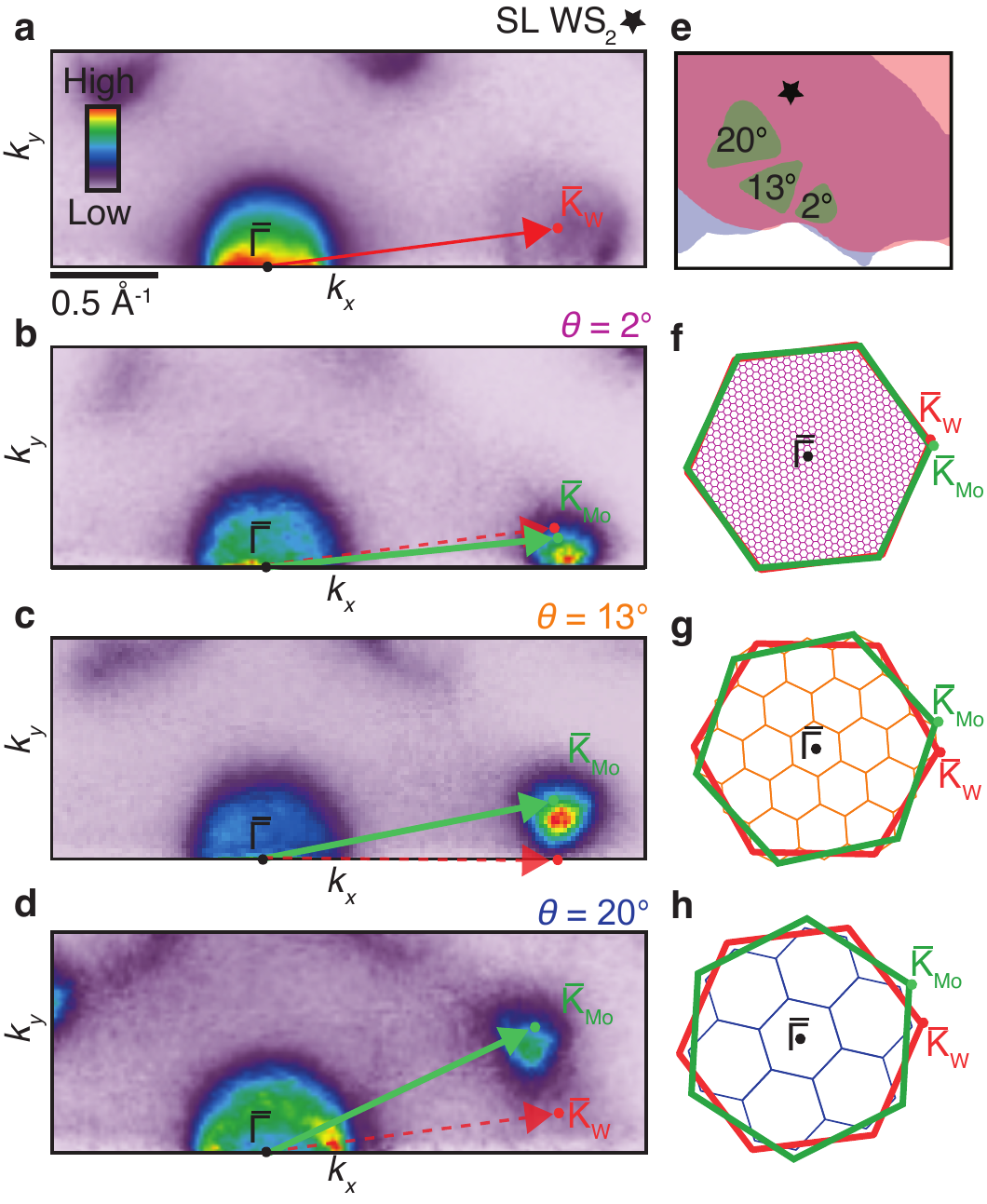}
		\caption{Twist angles and moir\'e mBZs: (a)-(d) $(k_x,k_y)$-dependent constant energy surfaces at -2~eV for (a) SL WS$_2$ and (b)-(d) MoS$_2$/WS$_2$ heterostructures with the given twist angles $\theta$. The red (green) arrows connect $\bar{\mathrm{\Gamma}}$ and $\bar{\mathrm{K}}_{\mathrm{W}}$ ($\bar{\mathrm{K}}_{\mathrm{Mo}}$) high symmetry points in the SL WS$_2$ (MoS$_2$/WS$_2$) constant energy surface. The dashed red arrows in (b)-(d) are based on the orientation of the high symmetry points in the WS$_2$ immediately adjacent to the MoS$_2$ islands. (e) Updated diagram of the sample composition from Figure \ref{fig:1}(c) with twist angles determined from (b)-(d) annotated on the corresponding MoS$_2$ islands. The constant energy surface in (a) was obtained on bare WS$_2$ from the area marked by a star in (e). (f)-(h) Relative BZ orientations of WS$_2$ (red hexagons) and MoS$_2$ (green hexagons) for each twist angle. The moir\'e mBZs have been drawn to fill out the main BZs to give a sense of reciprocal lattice length scales in each case. The ARPES spectrum in (a) was measured at 145 eV, while those in (b) to (d) were measured at 126 eV.}
		\label{fig:3}
	\end{center}
\end{figure*}

With the individual heterostructures spatially outlined, we are now able to single-out the full $E(k_x,k_y)$-dispersion for each heterostructure in order to extract the twist angle, $\theta$. Figures \ref{fig:3}(a)-(d) display $(k_x,k_y)$-dependent constant energy surfaces extracted at -2 eV for areas that contain the (a) SL WS$_2$ and (b)-(d) three MoS$_2$/WS$_2$ heterostructures observed in Figure \ref{fig:2}(e). The constant energy cuts provide access to band contours around $\bar{\mathrm{\Gamma}}$, $\bar{\mathrm{K}}_{\mathrm{W}}$ and $\bar{\mathrm{K}}_{\mathrm{Mo}}$. The constant energy surface of SL WS$_2$ in Figure \ref{fig:3}(a) displays an intense half-circle, which is identified as the hole-like band around $\bar{\mathrm{\Gamma}}$, as seen via the $E(k)$-plot in Figure \ref{fig:1}(e). Three surrounding triangular pockets are visible within the probed region of $k$-space, which originate from the $\bar{\mathrm{K}}_{\mathrm{W}}$-points. The main triangular pocket within the viewed region displays two concentric contours, which correspond to the spin-orbit split branches that are seen in the $E(k)$-plot in Figure \ref{fig:1}(f). By locating the centre of the pocket we determine the vector connecting $\bar{\mathrm{\Gamma}}$ and $\bar{\mathrm{K}}_{\mathrm{W}}$, which is shown as a red arrow in Figure \ref{fig:3}(a). Repeating this analysis for the three MoS$_2$/WS$_2$ heterostructures using the contours around the $\bar{\mathrm{K}}_{\mathrm{Mo}}$-points leads to the vectors connecting $\bar{\mathrm{\Gamma}}$ and $\bar{\mathrm{K}}_{\mathrm{Mo}}$ for each heterostructure, as seen via green arrows in Figures \ref{fig:3}(b)-(d). Note that the centres of the pockets are not necessarily determined by where they display a maximum in intensity, but is based on a visual inspection of how the contours develop with energy in the full $(E,k_x,k_y)$-dependent intensity, as is shown in Supporting Figure 10. Dashed red arrows then denote the $\bar{\mathrm{\Gamma}}-\bar{\mathrm{K}}_\mathrm{_W}$ direction on these heterostructures. Note that the WS$_2$ flake is observed to be slightly rotated between the heterostructures, which we account for by determining the orientation of the WS$_2$ band contours in areas immediately adjacent to each MoS$_2$ island. The angles between the green and red vectors correspond to the twist angles, which we determine as $(2.0 \pm 0.5)^{\circ}$, $(13 \pm 0.5)^{\circ}$, and $(20 \pm 0.5)^{\circ}$ for the heterostructures in Figure \ref{fig:3}(b), \ref{fig:3}(c), and \ref{fig:3}(d), respectively. Figure \ref{fig:3}(e) provides an updated version of the diagram from Figure \ref{fig:1}(c) with twist angles annotated for each heterostructure. 

Based on the extracted values of $\theta$ we calculate the length of the moir\'e reciprocal lattice vector as $G_m=G(1-\cos\theta)/\cos\phi$, where $G = 2.3$~\AA$^{-1}$ is the reciprocal lattice parameter for SL MoS$_2$ and SL WS$_2$\cite{Wajabayashi1975_MoS2neutron,Schutte1987_WS2}, and $\phi=\arctan[\sin\theta/(\cos\theta-1)]$. We obtain $G_m =\{(0.08 \pm 0.03)~\mathrm{\AA}^{-1},(0.52 \pm 0.04)~\mathrm{\AA}^{-1},(0.80 \pm 0.04)~\mathrm{\AA}^{-1}\}$ for $\theta =\{(2.0 \pm 0.5)^{\circ},(13 \pm 0.5)^{\circ},(20 \pm 0.5)^{\circ}\}$ respectively. The corresponding errors are propagated from the uncertainty in determining the twist angle. The resulting BZ and mBZ constructions for each of the heterostructures are shown in Figures \ref{fig:3}(f)-(h), illustrating the wide range of superlattice sizes that are accessible within our sample. High symmetry points of higher order mBZs do not line up with the symmetry points of main BZs due to the incommensurate nature of our superlattices.

\begin{figure*} [t!]
	\begin{center}
		\includegraphics[width=1\textwidth]{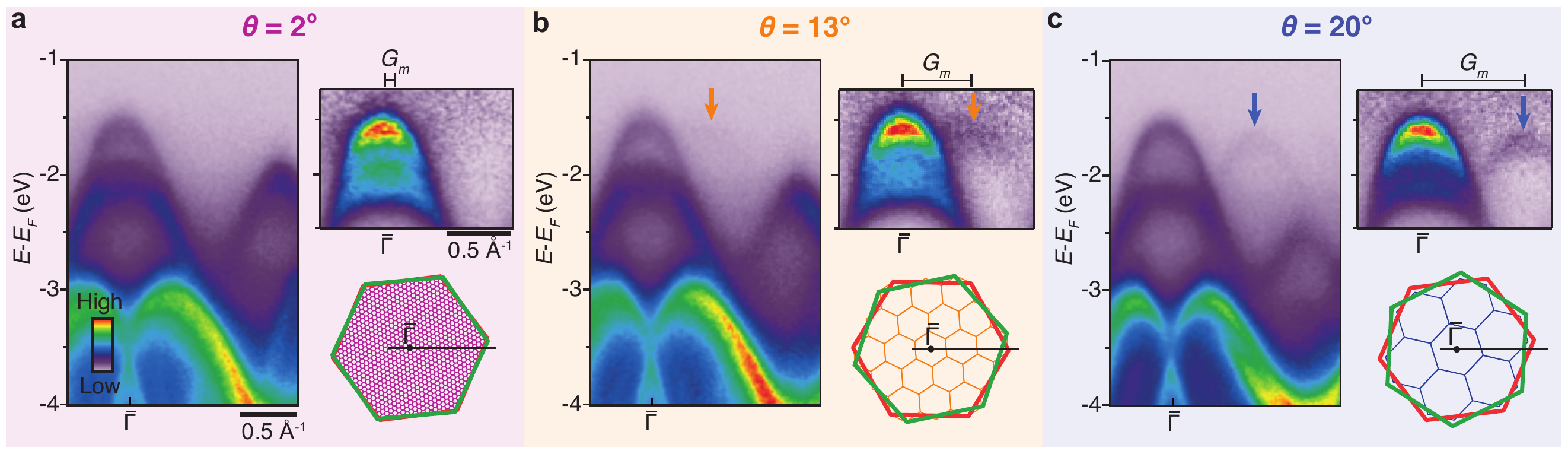}
		\caption{Dispersion of twisted MoS$_2$/WS$_2$ heterobilayers: (a)-(c) ARPES $E(k)$-dispersion with BZ and mBZ constructions for twist angles between MoS$_2$ and WS$_2$ given by (a) $\theta = 2^{\circ}$, (b) $\theta = 13^{\circ}$ and (c) $\theta = 20^{\circ}$. The black lines on the BZs indicate the cut direction for the ARPES spectra. The upper inserts with the dispersion present the top-most VB on an enhanced intensity scale with non-linear background normalization. The size of the corresponding moir\'e lattice vector $G_m$ is indicated via a scale bar above the inserts. Orange and blue arrows in (b)-(c) demarcate replica bands. The lower insets present the MoS$_2$ and WS$_2$ BZ orientations in green and red, respectively, with the measurement direction for each panel described with a black line. All ARPES spectra presented here were obtained using 126 eV.}
		\label{fig:4}
	\end{center}
\end{figure*}

The impact on the MoS$_2$/WS$_2$ $E(k)$-dispersion of varying $\theta$ is explored in Figure \ref{fig:4}. Here, we present the VB dispersion for the cuts through $\bar{\mathrm{\Gamma}}$ indicated in the BZ sketches in the corresponding panels in order to track replica bands throughout the mBZs. The overall VB features are very similar across all the heterostructures, although the dispersion away from $\bar{\mathrm{\Gamma}}$ varies with the different cuts through the main BZs. We provide EDCs through $\bar{\mathrm{\Gamma}}$ for each of the heterostructures in Supporting Information Figure 11, and find that the splitting at $\bar{\mathrm{\Gamma}}$ remains $\approx 550$~meV across the three heterostructures, within the error bars of our EDC analysis. In order to emphasize the location of the faint moir\'e replica bands, we have plotted the ARPES intensity around $\bar{\mathrm{\Gamma}}$ using an enhanced intensity scale with non-linear background normalization in the inserts. In the case of $\theta = 2^{\circ}$, shown in Figure \ref{fig:4}(a), no replicas are visible. The very small value of $G_m$ places low order replica states nearly on top of the intense main bands at $\bar{\mathrm{\Gamma}}$. For the larger twist angles of $\theta = 13^{\circ}$ and $\theta = 20^{\circ}$ in Figures \ref{fig:4}(b)-(c) the larger values of $G_m$ result in the low order replica states being $k$-shifted sufficiently far away from $\bar{\mathrm{\Gamma}}$ such that they can be distinguished from the main bands, as shown via the arrows in the dispersion plots. Note that the replica bands are displaced in energy with respect to the original bands due to the measurement not being through the high symmetry points of the mBZ, as is described in detail later. Higher order replica states are suppressed, likely due to incoherency between superlattice wavefunctions causing the reciprocal space Fourier components to rapidly diminish \cite{Ulstrup2020,Diaye2008}. 

\begin{figure*} [t!]
	\begin{center}
		\includegraphics[width=0.8\textwidth]{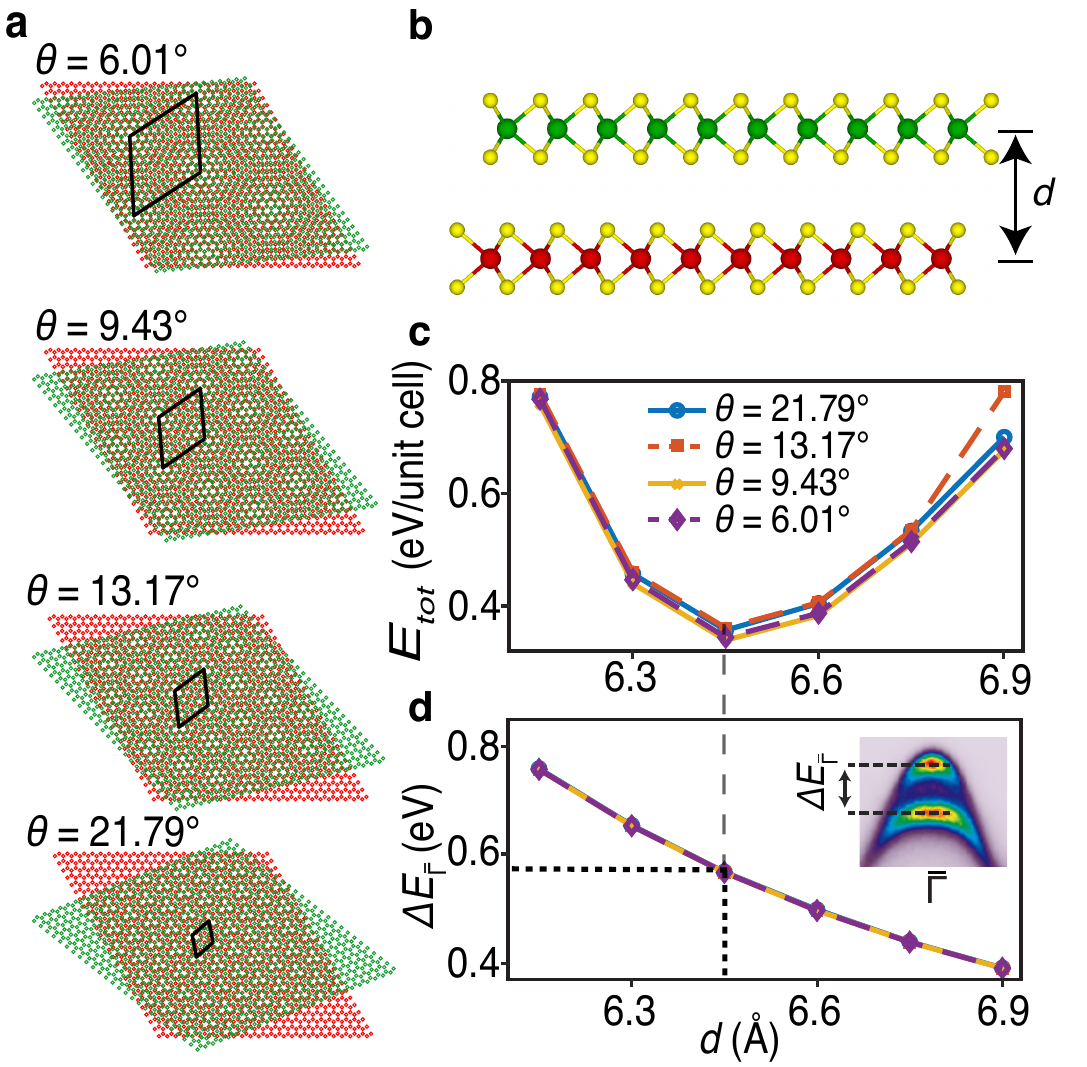}
		\caption{DFT calculations for twisted bilayer MoS$_2$/WS$_2$: (a) Real space MoS$_2$ (green) and WS$_2$ (red) lattices for four commensurate twist angles used in the calculations. The supercell is outlined in black. (b) Sketch of assumed 2H stacking with interlayer distance $d$. Mo, W and S atoms are color-coded green, red and yellow, respectively. (c) Total energy per unit cell as a function of interlayer distance for each twist angle. (d) Splitting at $\bar{\mathrm{\Gamma}}$ as a function of interlayer distance for each twist angle. Black dashed lines mark the calculated splitting corresponding to the optimum interlayer distance. The inset shows an example ARPES spectrum around $\bar{\mathrm{\Gamma}}$ with the splitting marked by a double-headed arrow and horizontal dashed lines.}
		\label{fig:5}
	\end{center}
\end{figure*}

We compare the electronic structure of our experimental incommensurate MoS$_2$/WS$_2$ heterostructures with DFT calculations based on commensurate supercells with twist angles given by $\theta =\{6.01^{\circ},9.43^{\circ},13.17^{\circ},21.79^{\circ}\}$, as shown in Figure \ref{fig:5}(a). These angles were selected as the closest commensurate structures to our incommensurate measured heterostructures, however smaller twist angles were not used due to the prohibitively large supercells required for calculation. The TMDs are assumed to stack in the 2H structural modification, as sketched in Figure \ref{fig:5}(b). In order to determine the interlayer distance, $d$, we perform a $\bar{\mathrm{\Gamma}}$-point total energy calculation with fixed in-plane positions while varying $d$. Details of the calculation can be found in the Methods section. In Figure \ref{fig:5}(c), we show that the total energy normalized by the number of unit cells in each supercell has a minimum at 6.45~\AA~and that this distance does not strongly depend on the twist angle. The splitting at $\bar{\mathrm{\Gamma}}$ is indicative of the degree of hybridization of the layers \cite{latini2017interlayer}, and indeed it strongly increases as the interlayer distance is reduced, as seen via the calculated values, $\Delta E_{\bar{\mathrm{\Gamma}}}$, in Figure \ref{fig:5}(d). However, since we find the structures relax to the same interlayer distance for the studied twist angles, and we see that the splitting appears insensitive to twist angle, the value of $\Delta E_{\bar{\mathrm{\Gamma}}}$ remains fixed around 550~meV for all calculated heterostructures, which is fully consistent with our experimental ARPES dispersion for the three heterostructures in Figure \ref{fig:4}.

\begin{figure*} [t!]
	\begin{center}
		\includegraphics[width=1\textwidth]{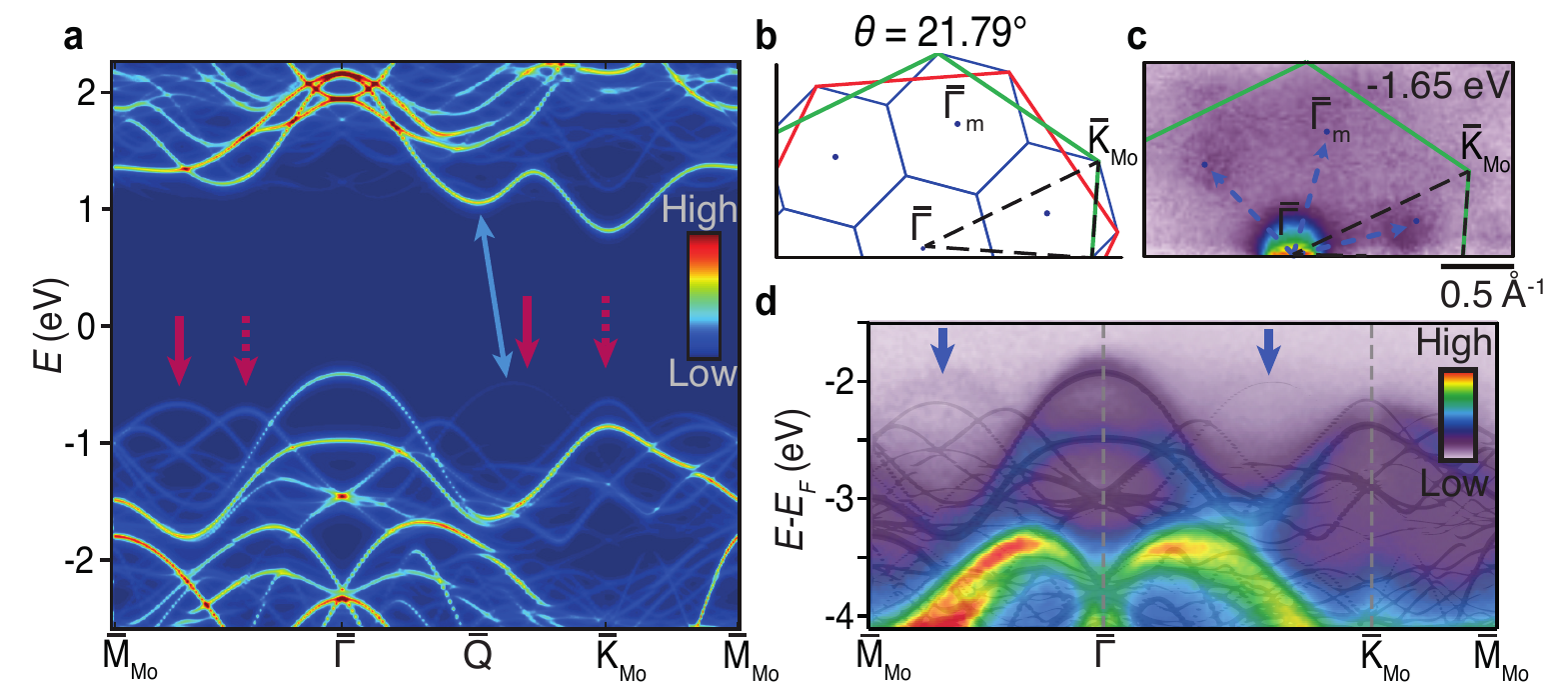}
		\caption{Superlattice dispersion from DFT and ARPES: (a) DFT calculation of MoS$_2$/WS$_2$ dispersion for  $\theta = 21.79^{\circ}$ unfolded along $\bar{\mathrm{M}}_{\mathrm{Mo}}-\bar{\mathrm{\Gamma}}-\bar{\mathrm{K}}_{\mathrm{Mo}}-\bar{\mathrm{M}}_{\mathrm{Mo}}$. Red arrows mark faint hole-like replicas arising from the moir\'e. A double-headed arrow marks a new indirect gap between a $\bar{\mathrm{Q}}$-valley of the CB and a replica band in the VB. The band intensity corresponds to the modulus square of the Fourier coefficients of the mBZ Bloch states evaluated on the MoS$_2$ BZ. (b) BZs for MoS$_2$ (green) and WS$_2$ (red) and commensurate mBZs for $\theta = 21.79^{\circ}$. The black triangular path marks the investigated high symmetry direction of the MoS$_2$ BZ. (c) ARPES constant energy surface at -1.65~eV from the 20$^{\circ}$-heterostructure. The MoS$_2$ BZ is outlined in green and moir\'e reciprocal lattice vectors are shown via blue dashed arrows. (d) ARPES dispersion along the path highlighted by a black triangle in (c). The unfolded DFT bands from (a) have been overlaid. Blue arrows mark replica bands in the ARPES dispersion. The overlaid DFT bands have been shifted rigidly in energy such that the VBM at $\bar{\mathrm{\Gamma}}$ overlaps that of the measurement.}
		\label{fig:6}
	\end{center}
\end{figure*}

We utilize the structure for the commensurate heterostructure with $\theta = 21.79^{\circ}$ to calculate the electronic bandstructure unfolded along the $\bar{\mathrm{M}}_{\mathrm{Mo}}-\bar{\mathrm{\Gamma}}-\bar{\mathrm{K}}_{\mathrm{Mo}}-\bar{\mathrm{M}}_{\mathrm{Mo}}$ high symmetry path of the MoS$_2$ BZ in order to compare the superlattice dispersion with our ARPES dispersion of the  20$^{\circ}$-heterostructure. We focus on the case of the larger mBZ here (i.e. the large twist angle), as the superlattice features here are most clearly defined in the experimental data. The calculated dispersion is presented in Figure \ref{fig:6}(a), where the intensity denotes the modulus square of the Fourier coefficients of the mBZ Bloch states evaluated on the MoS$_2$ BZ\cite{Mayo2020_unfoldedbands}. The calculation reveals an indirect bandgap between the VB maximum at $\bar{\mathrm{\Gamma}}$ and the conduction band (CB) minimum at $\bar{\mathrm{K}}_{\mathrm{Mo}}$. There are a number of faint hole-like bands, labelled with red arrows, arising from the moir\'e superstructure. The lower mass bands (dashed arrows) correspond to replicas of the moir\'e $\bar{\mathrm{K}}_{\mathrm{m}}$ points, and the higher mass bands (solid arrows) replicas of the VB maximum around the $\bar{\mathrm{\Gamma}}_{\mathrm{m}}$ points. Their energy maxima do not line up with the main bands, as the high symmetry cut of the main MoS$_2$ BZ does not traverse the moir\'e high symmetry points, which is seen in the black dashed path sketched in Figure \ref{fig:6}(b). We see that the unfolded moir\'e band structure gives rise to numerous new indirect transitions between local CB minima and the replica bands, for example the transition between a  $\bar{\mathrm{\Gamma}}_{\mathrm{m}}$ replica band and the CB minimum $\bar{\mathrm{Q}}$, labelled with a blue arrow. 

The unfolded bandstructure is compared with the ARPES dispersion of the 20$^{\circ}$- heterostructure. The constant energy surface at -1.65~eV in Figure \ref{fig:6}(c) outlines the centre of three mBZs within the probed region of $k$-space and indicates the main MoS$_2$ BZ. Each mBZ exhibits a pocket of  increased intensity, which derives from replicas of the VB maximum, in agreement with the sketch in Figure \ref{fig:6}(b). In Figure \ref{fig:6}(d), we present a cut through our data along the high symmetry path indicated in Figures \ref{fig:6}(b)-(c). Calculated bands in grey have been overlaid and rigidly shifted in energy to overlap the VB maximum at $\bar{\mathrm{\Gamma}}$. Several features are in agreement between DFT and ARPES data, including the hybridization at $\bar{\mathrm{\Gamma}}$ into a splitting of $\approx550$ meV, the replica bands near $\bar{\mathrm{\Gamma}}_m$ (marked with blue arrows), and the dispersion of lower VB subbands around -2~eV. We do not observe intensity arising from the $\bar{\mathrm{K}}_{\mathrm{m}}$ points, i.e. replicas of the bands around $\bar{\mathrm{K}}_{\mathrm{Mo}}$, which may be explained by the in-plane character of the orbitals associated with $\bar{\mathrm{K}}_{\mathrm{Mo}}$ leading to much weaker hybridization of these states between the TMDs \cite{kormanyos2015}. Hybridization is, however, seen at $\bar{\mathrm{\Gamma}}$, which leads to clear replica bands at $\bar{\mathrm{\Gamma}}_m$ in both the experimental data and the calculations. These results illustrate the significant effects of hybridization upon the electronic structure of our twisted heterostructures. The additional moir\'e bands open up the possibility of new indirect or optical transitions in the heterostructures, which may be highly tunable with twist angle, demonstrated by the new indirect transition arising between $\bar{\mathrm{Q}}$ and a $\bar{\mathrm{\Gamma}}_m$ replica band.

\section{Conclusion}

We have utilized microARPES to spatially resolve the electronic structure of SL MoS$_2$ islands stacked on SL WS$_2$ with twist angles of $(2.0 \pm 0.5)^{\circ}$, $(13.0 \pm 0.5)^{\circ}$, and $(20.0 \pm 0.5)^{\circ}$, leading to MoS$_2$/WS$_2$ heterobilayers supported on hBN. Our measurements reveal a significant hybridization effect between the TMDs, leading to a splitting of (548 $\pm$ 33)~meV around $\bar{\mathrm{\Gamma}}$ in addition to replica bands of these hybridized states at the expected moir\'e reciprocal lattice vectors. A comparison between ARPES dispersion and DFT calculations of commensurate MoS$_2$/WS$_2$ heterobilayers with similar twist angles confirms that the splitting around $\bar{\mathrm{\Gamma}}$ mainly depends on the MoS$_2$/WS$_2$ interlayer distance and not on twist angle, while a direct-to-indirect bandgap transition is observed from the bare SL WS$_2$ to the MoS$_2$/WS$_2$ heterobilayers. The additional moir\'e replicas induce $\theta$-dependent  transitions between CB and VB states, highlighting the possibility to engineer optoelectronic properties of TMD heterostructures using the twist angle-tunable superlattice dispersion.

\section{Materials and Methods}

\subsection{Preparation of heterostructures}

Heterostructure construction began by exfoliating hBN onto SiO$_2$(300\,nm)/Si and then transferring it to a 0.5 wt\% Nb-doped rutile TiO$_2$ (100) (Shinkosha Co., Ltd) using a thin film of polycarbonate (PC) laid onto a polydimethylsiloxane (PDMS) stamp.  Upon melting the PC, the hBN/TiO$_2$ was cleaned with chloroform and isopropyl alcohol (IPA). Both the WS$_2$ and MoS$_2$ monolayers were grown through chemical vapor deposition (CVD) in similar growth techniques to previous work\cite{Cunningham2016MoS2,Cunningham2016WS2}.  Using another PC transfer slide, these monolayers were successively picked up such that the WS$_2$ monolayer completely covered three MoS$_2$ islands. Finally, the TMD materials were “dropped off” by melting the PC on the TiO$_2$/hBN such that the islands were completely laid on top of the hBN.  The WS$_2$, being much larger than the MoS$_2$ and hBN, is in contact with the underlying TiO$_2$ substrate, allowing for a conducting pathway to re-fill the photoemitted states.  The sample was cleaned with chloroform and IPA and then annealed in UHV to remove any remaining residues prior to exposure to the synchrotron beam.

\subsection{microARPES}
The microARPES measurements were performed at the MAESTRO facility of the Advanced Light Source, Lawrence Berkeley National laboratory. Measurements were carried out using a Scienta R4000 hemispherical analyser with the entrance slit aligned parallel to the polar rotation axis of the chamber. The best energy and momentum resolutions achieved were 40 meV and 0.01~\AA$^{-1}$, respectively. All measurements were performed with the sample fixed at room temperature. 

The synchrotron beam was focused to a spot size of $(1.8 \pm 0.3)$~$\mu$m using a capillary X-ray optic from Sigray Inc. \cite{Koch2018,Ulstrup2020}. The spatially-dependent photoemission intensity data presented in Figures \ref{fig:1}-\ref{fig:2} were collected by scanning the beam $(x,y)$-position relative to the sample in 0.5~$\mu$m steps and measuring a photoemission spectrum at each position. The $(E,k_x,k_y)$-dependent data presented in Figures \ref{fig:3} and \ref{fig:6} were collected by first moving the sample $(x,y)$-position to the desired heterostructure or bare SL WS$_2$ area and then measuring ARPES spectra while rotating the detector polar angle, keeping the sample position and orientation fixed. The ARPES data presented in Figures \ref{fig:1} and \ref{fig:3}(a) were measured with a photon energy of 145 eV, all other ARPES data was measured using 126 eV photon energy. The core level data presented in Figure \ref{fig:2} were measured using a 350 eV photon energy for the S 2p and Mo 3d core levels, and 145 eV photon energy for the W 4f core levels. The energies of all bands were referenced to the Fermi level, which was calibrated using measurements on the polycrystalline sample holder. 

\subsection{DFT calculations}
The Vienna $Ab$ $initio$ Simulation Package (VASP)\cite{kresse1996efficiency,kresse1999ultrasoft} was used for DFT calculations. The exchange-correlation potentials were described through the Perdew-Burke-Ernzerhof (PBE) functional within the generalized gradient approximation (GGA) formalism \cite{perdew1996generalized}. A plane wave basis set was used with a cutoff energy of 300 eV. A vacuum region of 17~\AA~was set between the two heterostructures in order to minimize the interaction between the periodic repetitions of the cell. The zero damping DFT-D2 method of Grimme \cite{grimme2006semiempirical} was used to account for the long range vdW interaction between the monolayers. MoS$_2$ and WS$_2$ are lattice matched with the PBE relaxed in-plane lattice constant $a = 3.18$~\AA~to study the commensurate moir\'e superstructures corresponding to $\theta =\{6.01^{\circ},9.43^{\circ},13.17^{\circ},21.79^{\circ}\}$. $\bar{\mathrm{\Gamma}}$-point total energy calculations were performed with fixed in-plane parameters for the interlayer distances in the range of 6.15~\AA~to 6.90~\AA. We neglected the spin-orbit interaction in these calculations. It is expected that the WS$_2$ VB states at $\bar{\mathrm{\Gamma}}$ move to higher energies\cite{Zhu2011}, however, we ignore this effect in this work due to computational costs of large supercell calculations.
For the unfolded bands presented in Figure \ref{fig:6}, we adopted a $4\times4$ Monkhorst-Pack \cite{monkhorst1976special} $k$-point mesh and used the corresponding Bloch functions to unfold the bands on the BZ of the MoS$_2$. The intensity of DFT bands shown in Figures \ref{fig:6}(a) and \ref{fig:6}(d) was computed using $P(K,E) = \sum_{kn} |\langle K|u_{kn}\rangle|^2 \delta(E-\epsilon_{kn})$, where $\phi_{kn} = u_{kn} e^{ikx}$ are the mBZ Kohn-Sham Bloch states of the heterostructure, $\epsilon_{kn}$ are the corresponding energies and $|K\rangle$ is a plane wave evaluated at a momentum $K$ belonging to the MoS$_2$ BZ.\cite{Mayo2020_unfoldedbands}

\begin{acknowledgement}
S. U. acknowledges financial support from the Independent Research Fund Denmark under the Sapere Aude program (Grant No. 9064-00057B) and VILLUM FONDEN under the Young Investigator Program (Grant No. 15375). J. K. acknowledges the financial support from U.S. Department of Energy, Office of Science, Office of Basic Energy Sciences, under Award Number DE-SC0020323. J. A. M. acknowledges financial support from the Independent Research Fund Denmark under the Sapere Aude program (Grant No. DFF-6108-00409). The authors also acknowledge the Villum Centre of Excellence for Dirac Materials (Grant No. 11744). The Advanced Light Source is supported by the Director, Office of Science, Office of Basic Energy Sciences, of the U.S. Department of Energy under Contract No. DE-AC02-05CH11231. The work at NRL was supported by core programs and the Nanoscience Institute. Growth of hexagonal boron nitride crystals was supported by the Elemental Strategy Initiative conducted by the MEXT, Japan, Grant Number JPMXP0112101001 and  JSPS KAKENHI Grant Number JP20H00354.
\end{acknowledgement}

\section{Notes}

Correspondence and requests for materials should be addressed to J. K. (jkatoch@andrew.cmu.edu) or S. U. (ulstrup@phys.au.dk).\\
\\
The authors declare that they have no competing financial interests.

\providecommand{\latin}[1]{#1}
\makeatletter
\providecommand{\doi}
  {\begingroup\let\do\@makeother\dospecials
  \catcode`\{=1 \catcode`\}=2 \doi@aux}
\providecommand{\doi@aux}[1]{\endgroup\texttt{#1}}
\makeatother
\providecommand*\mcitethebibliography{\thebibliography}
\csname @ifundefined\endcsname{endmcitethebibliography}
  {\let\endmcitethebibliography\endthebibliography}{}

\newpage
\noindent\textbf{Supporting Information 1. EDC fitting of ARPES spectra}

\begin{figure*} [h!]
	\begin{center}
		\includegraphics[width=1\textwidth]{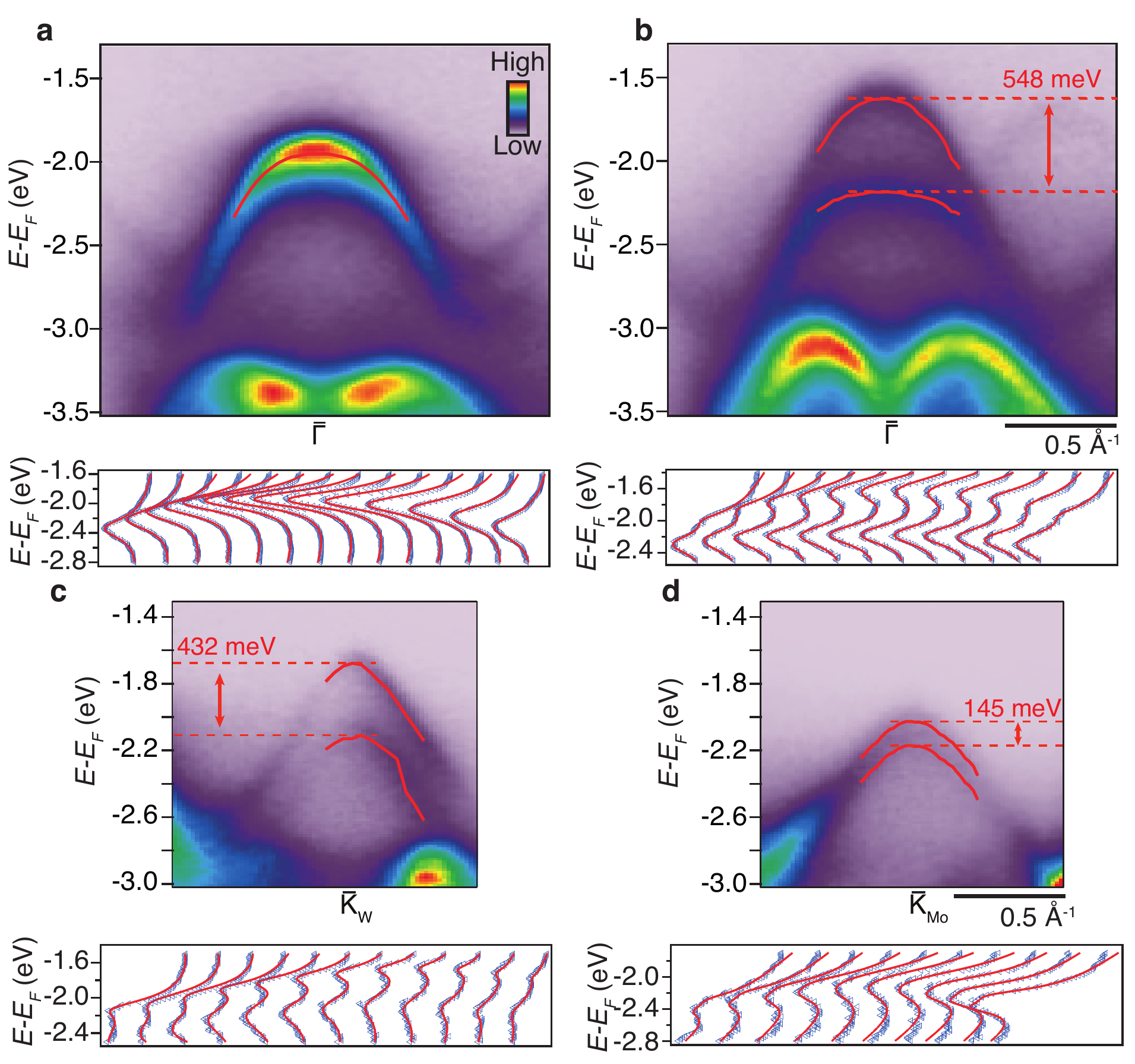}
		\caption{Energy distribution curve (EDC) analysis of ARPES spectra presented in Figure 1 of the main text. (a) Top panel: ARPES spectrum from SL WS$_2$ at $\bar{\mathrm{\Gamma}}$. The valence band (VB) has been fit to a Lorentzian and the fitted peak position overlaid as a red line. Measured splittings at $\bar{\mathrm{\Gamma}}$ and $\bar{\mathrm{K}}$ have been displayed with horizontal dashed lines. Bottom panel: Example EDCs around the VB (blue triangular markers) with individual fits overlaid (red curves). (b)-(d) Similar fits as shown in (a) for (b) the $\bar{\mathrm{\Gamma}}$ point of the MoS$_2$/WS$_2$ heterostructure, (c) the $\bar{\mathrm{K}}_{\mathrm{W}}$ point of SL WS$_2$, and (d) the $\bar{\mathrm{K}}_{\mathrm{Mo}}$ point of the heterostructure. For (d), the fitting of the two peaks was constrained using a constant separation of 145 meV.}
		\label{fig:S1}
	\end{center}
\end{figure*}

\newpage

\noindent\textbf{Supporting Information 2: Degeneracy of moir\'e replica bands and MoS$_2$/WS$_2$ states at $\bar{\mathrm{\Gamma}}$}

\begin{figure*} [h!]
	\begin{center}
		\includegraphics[width=1\textwidth]{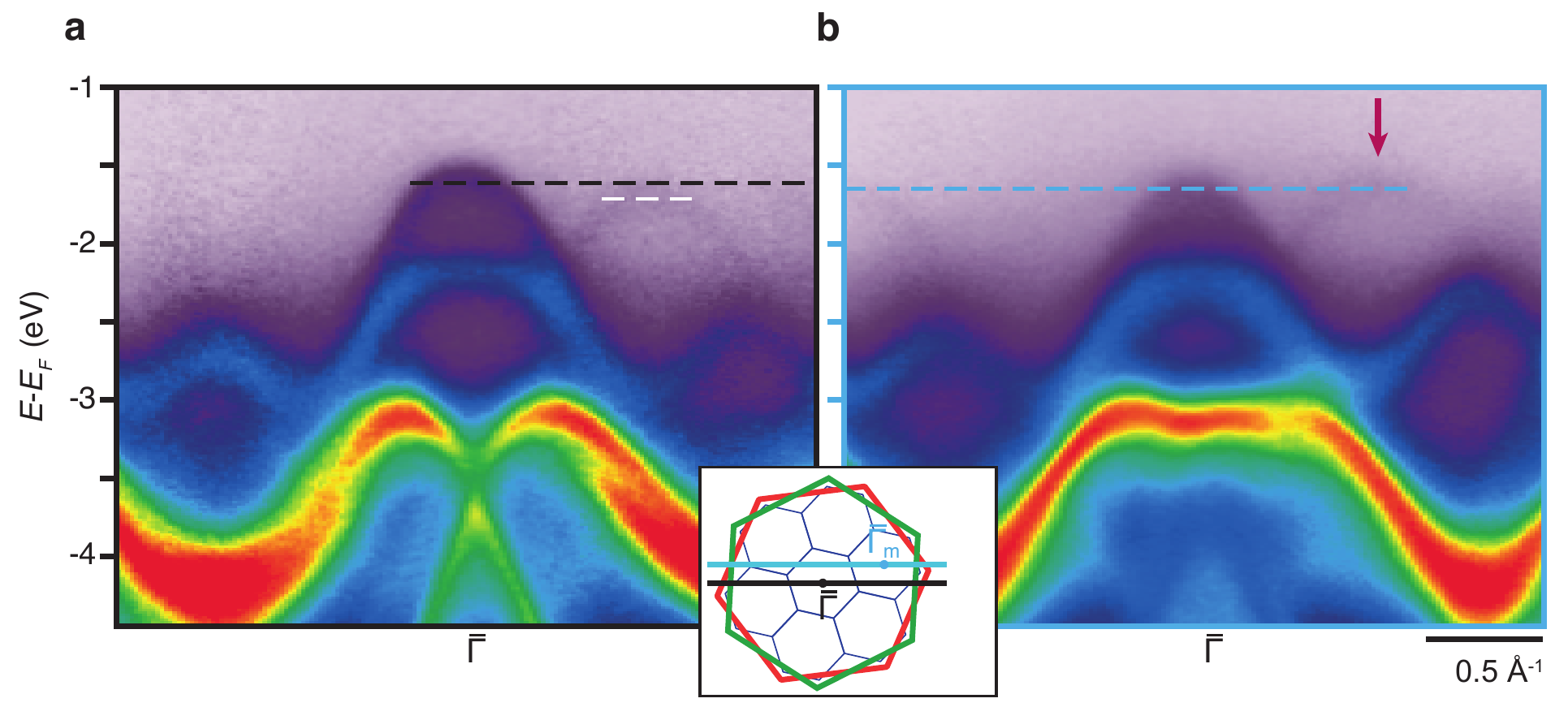}
		\caption{ARPES snapshots from a 20$^{\circ}$ twisted MoS$_2$/WS$_2$ heterostructure. (a) $(E,k)$ cut through the $\bar{\mathrm{\Gamma}}$ point of the MoS$_2$ BZ. A dashed black line marks the maximum of the VB at $\bar{\mathrm{\Gamma}}$, while a white dashed line labels the maximum of the replica band in this cut. (b) $(E,k)$ cut through the $\bar{\mathrm{\Gamma}}_m$ point of a mBZ. A dashed blue line marks the maximum of the replica band. The inset shows the BZ arrangement with the $k$-direction for the two ARPES spectra marked by solid lines.}
		\label{fig:S2}
	\end{center}
\end{figure*}

\newpage

\noindent\textbf{Supporting Information 3: Fitting parameters for core level analysis}

\begin{table}[h!]
	\begin{tabular}{|c|c|cc|cc|cc|}
		\hline
		\multicolumn{2}{|c|}{\cellcolor[HTML]{333333}{\color[HTML]{FFFFFF} Peak Name}} & \multicolumn{2}{c|}{\cellcolor[HTML]{333333}{\color[HTML]{FFFFFF} Binding Energy (eV)}}                     & \multicolumn{2}{c|}{\cellcolor[HTML]{333333}{\color[HTML]{FFFFFF} Lorentzian FWHM (eV)}} & \multicolumn{2}{c|}{\cellcolor[HTML]{333333}{\color[HTML]{FFFFFF} Gaussian FWHM (eV)}} \\ \hline
		\multicolumn{2}{|c|}{}& {\cellcolor[HTML]{C0C0C0}Peak 1} & \cellcolor[HTML]{C0C0C0}Peak 2  & {\cellcolor[HTML]{C0C0C0}Peak 1}   & \cellcolor[HTML]{C0C0C0}Peak 2   & \cellcolor[HTML]{C0C0C0}Peak 1   & {\cellcolor[HTML]{C0C0C0}Peak 2}  \\ \hline
		Mo 3d & MoS$_2$& 225.08 & \cellcolor[HTML]{FFFFFF}{\color[HTML]{333333} 228.17} & \cellcolor[HTML]{FFFFFF}{\color[HTML]{333333} 0.26}   & 0.17& 0.48& 0.68  \\
		& WS$_2$& 32.19 & 34.28  & 0.12 & 0.15  & 0.37  & 0.36   \\
		\multirow{-2}{*}{W 4f} & MoS$_2$& 32.17 & 34.27& 0.10 & -0.05  & 0.38 & 0.49  \\
		& WS$_2$  & 158.10   & 159.26& 0.18 & -0.06 & 0.33  & 0.39 \\
		\multirow{-2}{*}{S 2p}& MoS$_2$&158.00 & 159.16 & 0.20 & 0.24& 0.39& 0.35 \\ \hline                                               
	\end{tabular}
	\caption{Fitting parameters for the core level data presented in Figure 2 of the main paper. Fits were made to pairs of Voigt profiles to describe the core levels. The binding energy, Lorentzian and Gaussian widths from these fits are presented in eV in the table.}
	\label{tab:1}
\end{table}

\begin{figure*} [h!]
	\begin{center}
		\includegraphics[width=1\textwidth]{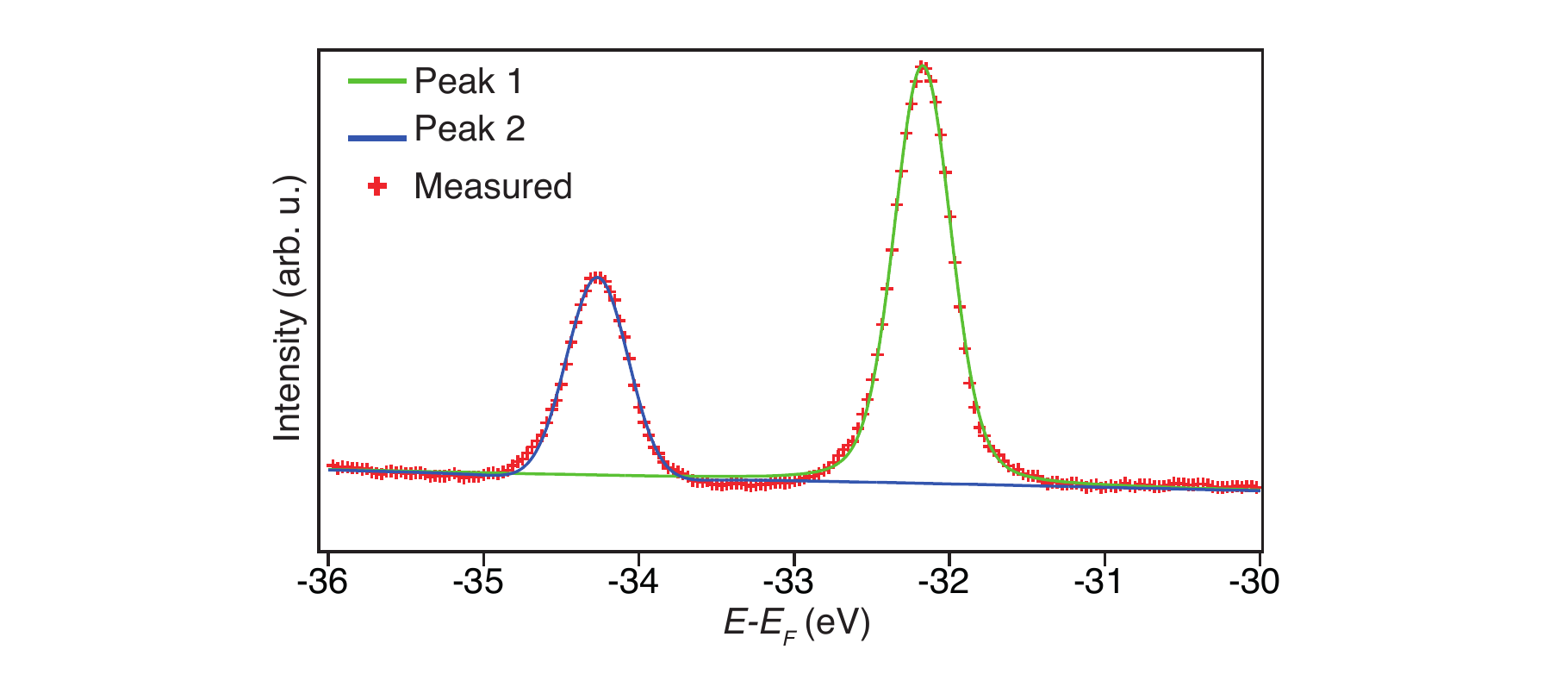}
		\caption{Example fitted core level spectrum from the W 4f core levels. The measured data points are marked with red crosses, and the individual Voigt peaks with solid coloured lines.}
		\label{fig:S3}
	\end{center}
\end{figure*}

\newpage

\noindent\textbf{Supporting Information 4: Evolution of valence band pockets}

\begin{figure*} [h!]
	\begin{center}
		\includegraphics[width=1\textwidth]{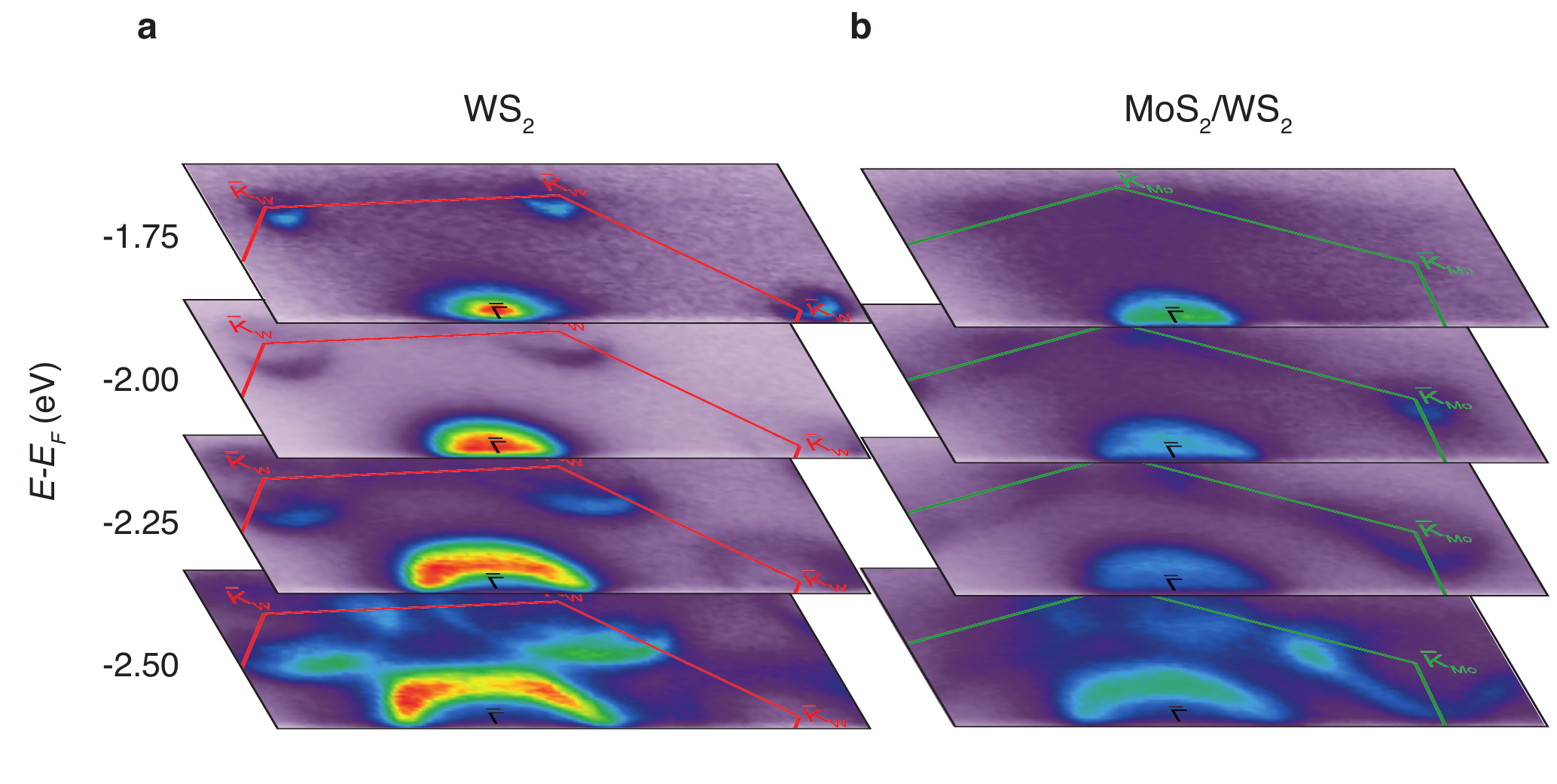}
		\caption{Constant energy surfaces between -2.50 and -1.75 eV from (a) WS$_2$ and (b) MoS$_2$/WS$_2$ regions. The extent of the WS$_2$ and MoS$_2$ BZs have been marked with coloured hexagons. }
		\label{fig:S4}
	\end{center}
\end{figure*}

\newpage
\noindent\textbf{Supporting Information 5: EDCs showing band splitting at $\bar{\mathrm{\Gamma}}$}

\begin{figure*} [h!]
	\begin{center}
		\includegraphics[width=1\textwidth]{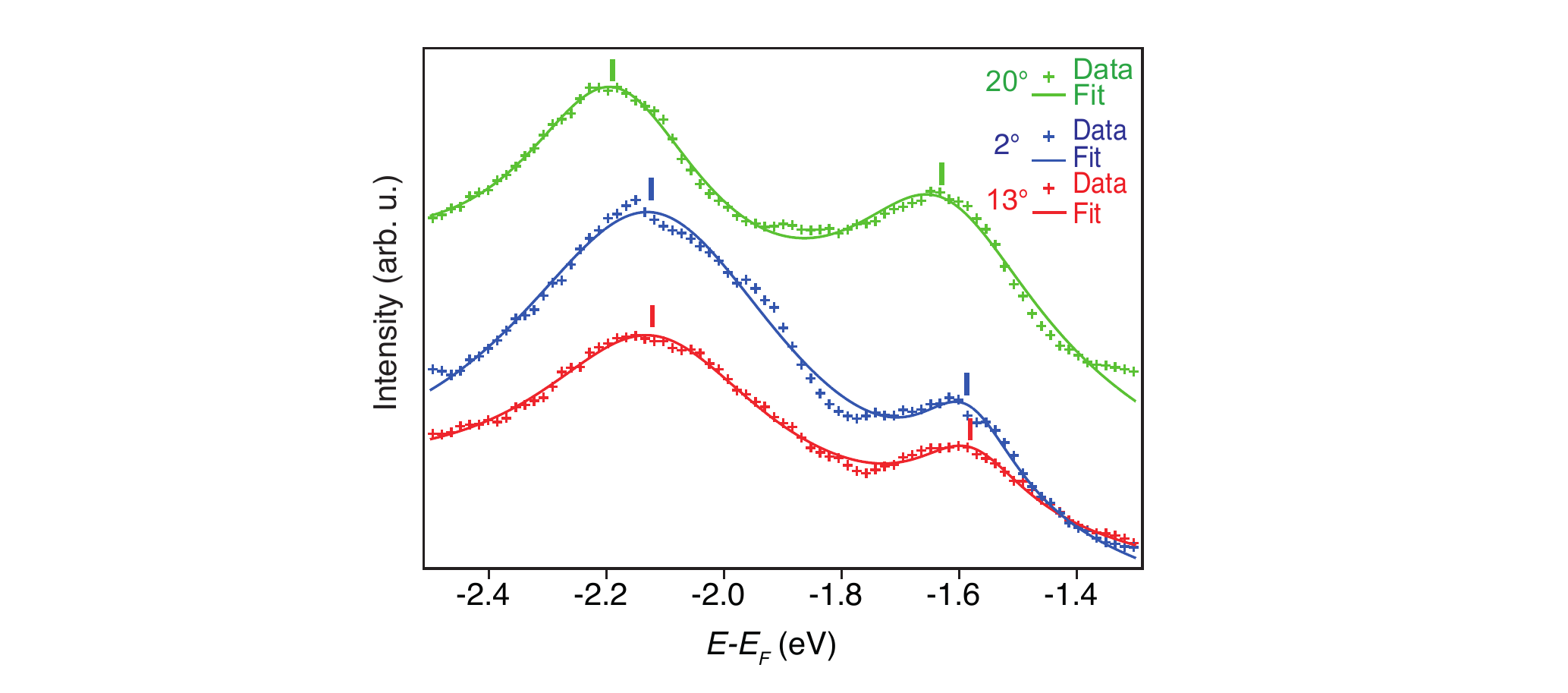}
		\caption{Measured energy distribution curves (crosses) and double Lorentzian fits (curves) through the $\bar{\mathrm{\Gamma}}$ point for the heterostructures with twist angles of 2$^{\circ}$, 13$^{\circ}$, and 20$^{\circ}$. Tick marks indicate the fitted peak positions. }
		\label{fig:S5}
	\end{center}
\end{figure*}

\end{document}